\documentclass[aps,pre,onecolumn,nofootinbib]{revtex4-2}
    \RequirePackage[sort&compress]{natbib}
    ﻿
    ﻿
    ﻿
    ﻿
    \usepackage{amsmath,amsfonts,amssymb}
    \usepackage{bm}
    \usepackage{physics}
    \usepackage{xcolor}
    \usepackage{graphicx}
    \usepackage{float}
    \usepackage{mathtools}
    \usepackage{tabularx}
    \usepackage{xr-hyper} 
    \usepackage{lpic,tikz}
    \usepackage[utf8]{inputenc}
    \usepackage[normalem]{ulem} 
    \usepackage{lineno}
    ﻿
    ﻿
    ﻿
    ﻿
    ﻿
    ﻿
    ﻿

    ﻿

    ﻿
    
    \definecolor{amethyst}{rgb}{0.6, 0.4, 0.85}
    \definecolor{darkgreen}{RGB}{0, 100, 0}
    ﻿
    
    ﻿
    
    ﻿
    ﻿
    ﻿
    ﻿
    ﻿
    ﻿
    ﻿
    \begin{document}
    ﻿
    \title{Network-Driven Vaccination Strategies for Preventing Rebound Dynamics in Metapopulation Epidemic Control}
    ﻿
    ﻿
    \author{
    Piergiorgio Castioni$^{1}$, Alex Arenas$^{1,2}$}
    ﻿
    \affiliation{$^{1}$Departament d'Enginyeria Inform\`{a}tica i Matem\`{a}tiques, Universitat Rovira i Virgili, 43007 Tarragona, Spain\\
    $^{2}$Pacific Northwest National Laboratory, 902 Battelle Boulevard, Richland, Washington 99354, USA}
    ﻿
    ﻿
    ﻿
    ﻿
    ﻿
    ﻿
    \begin{abstract}
    A critical question in epidemic control concerns the minimal requirements for a vaccination campaign to effectively halt a contagion process. However, control measures can inadvertently trigger resurgence dynamics, driven by a reservoir of susceptible individuals left unexposed in the controlled wave. This phenomenon, known as the “rebound effect”, is often preceded by a temporary drop in cases, termed usually as the “honeymoon period”. In this study, we examine the fundamental conditions for rebound dynamics within a metapopulation network framework. By elucidating the mechanisms underlying rebound events, we derive a rigorous mathematical criterion that identifies, based solely on the metapopulation network structure, the specific vaccination strategies likely to precipitate a rebound. Additionally, we propose an alternative vaccination protocol designed to eliminate rebound dynamics entirely. This approach is analytically validated and offers a robust pathway toward sustainable epidemic control.
    \end{abstract}
    ﻿
    ﻿
    ﻿
    \maketitle
    ﻿
    \section*{Introduction}
    Since the birth of the field of epidemic modeling, the subject of how to control and reduce the negative effects of an epidemic has been of paramount importance for researchers. In fact, the aim of the article that for the first time applied mathematics to the study of an epidemic, published by Daniel Bernoulli in 1766, was to quantify the number of lives saved from smallpox through extensive inoculation campaigns \cite{bernoulli_attempt_2004}. Even nowadays that same question remains extremely relevant, and in recent years we have seen a flourishing of papers dedicated to optimizing control interventions (both pharmaceutical \cite{klepac_synthesizing_2011, lauro_optimal_2021} or non-pharmaceutical \cite{wallinga_optimizing_2010, morris_optimal_2021,bubar_21}), where the criterion to define optimality heavily depends on the model and on the choice of the cost function. \\
    ﻿
    It has been noticed that interventions aimed at controlling the epidemic can themselves lead to unexpected negative effects once they are lifted \cite{pulsating_20,castioni_rebound_2024}. That is easy to see in the case of lockdown policies and risk perceptions, which lead to an increase in the number of contacts among susceptible people and therefore an increase in the reproduction number \cite{behavioural_22,protective_22}. However it has been shown that vaccination campaign can also produce similar results, although the reasons behind it are harder to understand. Even in the case of vaccines with life-long immunity, like the ones against rubella or measles, mass vaccination campaigns might result in an initially calm period (sometimes named the ``honeymoon period'' \cite{mclean_measles_1988, honey_21}) followed by an delayed but sudden resurgence, which usually affects communities that would have been left alone in the pre-vaccination era. Such an effect is usually connected to a slow build-up of the susceptible population, which can only happen because the number of infections is kept artificially low by the vaccine \cite{anderson_vaccination_1983, mclean_after_1995, scherer_mathematical_2002, bhattacharyya_age-specific_2017, munday_estimating_2024}. The same basic mechanism underlie the so-called ``rebound'' effect, where the interruption of the usage of a short-lasting immunity vaccine (such as the one for the flu or COVID19) can produce epidemic peak larger than the ones there would have been without the vaccination, depending critically on the campaign timing. \cite{castioni_rebound_2024}. \\ 
    This paper focuses on this effect, building on the previous work by Castioni et al. \cite{castioni_rebound_2024} to extend the criterion for epidemic rebound within the framework of network epidemiology. We will address a longstanding question in epidemic control, first posed in the 'city and villages' model \cite{may_spatial_1984}: is it more effective to prioritize vaccination for highly connected patches or for peripheral ones? Although results obtained in the mean-field scenario are not directly applicable to networked populations, some theoretical insights offer valuable perspectives for understanding the mechanisms of rebound dynamics in metapopulations and for identifying the optimal strategy to counteract them. \\
    ﻿
    The paper is structured as follows: First, we describe the model and the type of vaccination campaign under consideration. Second, we present a general criterion for determining whether a vaccination campaign may lead to a rebound. Third, we examine the relationship between vaccination coverage (i.e., the percentage of vaccinated patches) and the strength of network links, which in a metapopulation model typically corresponds to mobility between patches. Finally, we discuss a potential approach to eliminate the rebound effect without compromising the full utilization of available vaccines. 
    ﻿
    \begin{figure*}
        \centering
        \includegraphics[width=0.8\linewidth]{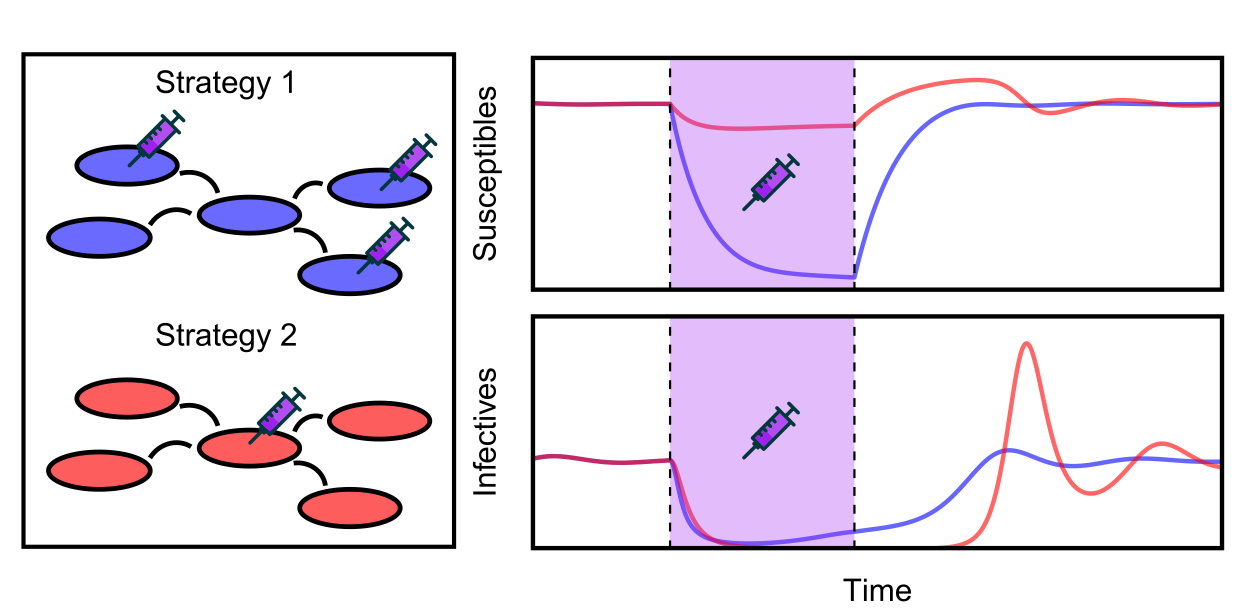}
        \caption{{\textbf{Rebound caused by different vaccination strategies in a star network}. In this example, two vaccination strategies are used: the first (blue) targets only the ``leaves'' of the graph, while the second (red) focuses solely on the hub. The vaccination campaign starts once equilibrium has been reached, as indicated by the shaded area and the syringe symbols. It can be clearly seen that, although strategy 1 employs more resources and lowers the susceptible population significantly, it does not result in a rebound because the infectious population never drops to zero. On the other hand, strategy 2 is much more effective at eradicating the disease across the whole network, thereby creating conditions for the susceptible population to grow above its equilibrium value, resulting in a noticeable rebound.}}
        \label{fig:scheme}
    \end{figure*}
    ﻿
    \subsection*{The model}
    We consider a continuous-time epidemic model within a cross-coupled metapopulation framework, where the state variables are represented by the vectors $S_i$, $I_i$, and $ R_i$. These vectors correspond to the fractions of susceptible, infectious, and recovered individuals, respectively, in each patch $i$, satisfying the normalization condition $S_i + I_i + R_i = 1$ for each patch. This condition ensures that the population fraction remains constant across each patch in the model.
    The dynamical equations of our SIRS model are the following:
    \begin{eqnarray} \label{eq:SIRS1}
    \dv{S_i}{t} &=& - \beta	S_i I_i - p \beta S_i \sum_{j=1}^{M} A_{ij} I_j + \delta R_i - \alpha_i(t), \\ \label{eq:SIRS2}
    \dv{I_i}{t} &=& \beta	S_i I_i + p \beta S_i \sum_{j=1}^{M} A_{ij} I_j - \mu I_i ,\\ \label{eq:SIRS3}
    \dv{R_i}{t} &=& \mu I_i - \delta R_i + \alpha_i(t)
    \end{eqnarray}
    where $\beta$ is the local transmission rate, $\mu$ is the recovery rate, $\delta$ is the waning immunity rate, $A_{ij}$ is the adjacency matrix of the contact network among patches (often referred to as the ``who acquires infection from whom'' WAIFW matrix; see \cite{sattenspiel_geographic_2009}), and $p \in [0,1]$ is the weight associated with each connection, representing the fraction by which the transmission rate is reduced if an infectious contact originates from a different patch. In the metapopulation context, $A_{ij}$ typically represents the mobility network, while $p$ serves as a factor to modulate mobility reductions. 
    ﻿
    Throughout this paper, Eqs.~\eqref{eq:SIRS1}-\eqref{eq:SIRS3} are always solved with the same initial conditions: $S_i = 0.99$ and $I_i = 0.01$, chosen to be identical for each patch to avoid introducing confounding sources of heterogeneity in the final results. 
    ﻿
    The time-dependent vaccination rate, $\alpha(t)$, is assumed to be non-zero only between times $t_{\mbox{\scriptsize start}}$ and $t_{\mbox{\scriptsize stop}}$:
    \begin{equation}
        \alpha(t) = \begin{cases}
            \alpha, & \text{if $t \in [t_{\mbox{\scriptsize start}}, t_{\mbox{\scriptsize stop}}]$} \\
            0, & \text{otherwise}.
        \end{cases}
    \end{equation}
    ﻿
    It should be noted that the basic reproduction number for this model depends on the network topology as follows:
    \begin{equation} \label{eq:R0(A)}
        R_0(\mathbf{A}) = \dfrac{\beta}{\mu} \left(1 + p \lambda_{\mbox{\scriptsize max}}(\mathbf{A}) \right),
    \end{equation}
    where $\lambda_{\mbox{\scriptsize max}}\bf(A)$ is the spectral radius of the adjacency matrix $\bf{A}$ \cite{granell_24}. This expression is significant because it demonstrates that an epidemic can spread even when the mean-field reproduction number, $\beta / \mu$, within a single patch is less than one. This feature is crucial for understanding whether a rebound effect may occur. 
    ﻿
    In a previous study \cite{castioni_rebound_2024}, vaccination rates were found to produce three distinct regimes: the coexistence regime, the eradication regime, and the total immunity regime. In the coexistence regime, defined by $\alpha \in [0, \alpha_{\text{er}}]$, all compartments (susceptible, infected, and recovered) are present, and the system’s behavior qualitatively resembles that of the standard SIR model, but with modified stationary values. The critical vaccination rate for eradication, $\alpha_{\text{er}} = (1 - \mu/\beta)\delta$, marks the threshold at which the endemic fraction of infected individuals drops to zero, subsequently allowing the susceptible population to build up sufficiently to fuel a new wave of infections once the vaccination campaign is halted.
    ﻿
    As a consequence, note that, in the mean-field case, the criterion for achieving such a rebound depends solely on the intensity of the vaccination campaign, with a critical value given by $\alpha_{\text{er}} = \delta (1 - \mu/\beta)$. However, in the case of the network model \eqref{eq:SIRS1}-\eqref{eq:SIRS3}, the spatial distribution of vaccines across population patches also plays an important role, for instance, by preventing localized extinctions of cases through colonization from neighboring patches. To isolate the specific contribution of this spatial aspect of the vaccination campaign, we assume that whenever vaccination is activated, $\alpha > \alpha_{\text{er}}$. That is, we investigate the role of metapopulation vaccination under the assumption that the vaccination rate is always sufficient to eradicate the infection at the level of the single patch.
     \\
    ﻿
    ﻿
    \subsection*{The rebound criterion}
    \begin{figure*}[b]
        \centering
        \includegraphics[width=0.8\linewidth]{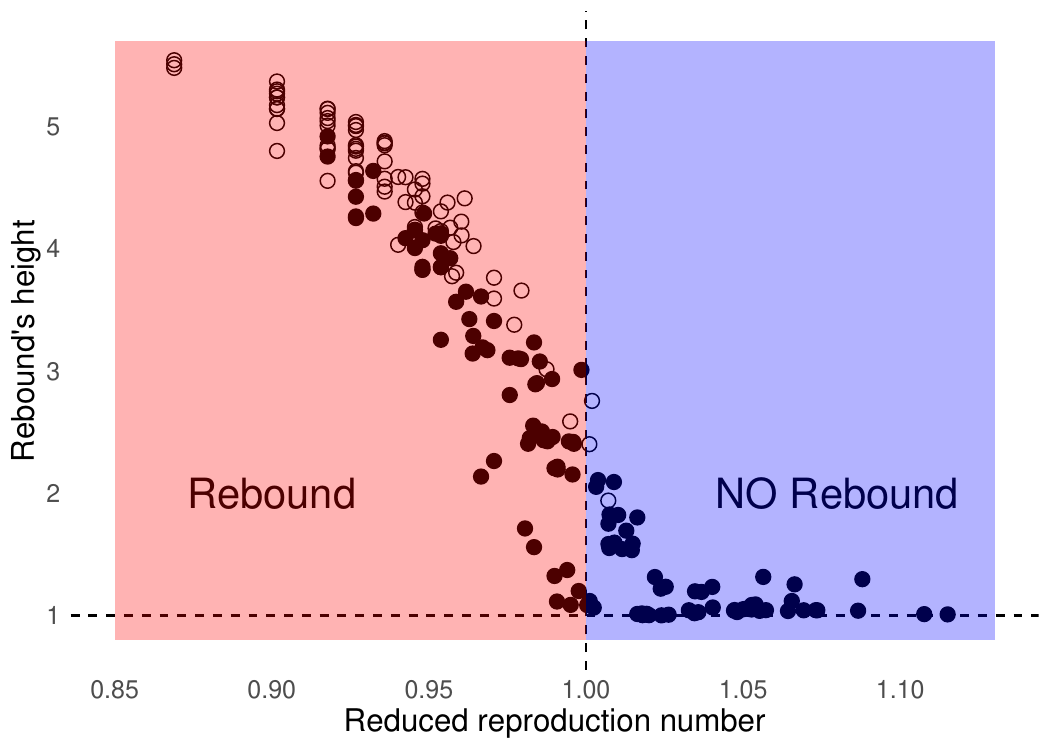}
        \caption{\textbf{Rebound in random vaccination campaigns.} The figure shows 200 different realizations of a vaccination campaign in which 160 patches out of 200, in a heterogeneous network with a power-law degree distribution generated using the Barabási-Albert (BA) model, were vaccinated. The rebound height is measured relative to the equilibrium value of the infected compartment, so the horizontal line corresponds to a case with no rebound at all. The empty dots represent realizations in which the top four highest-degree patches are among the vaccinated. It can be clearly seen that the reduced reproduction number, $R_0{\bf(A_{{\mbox{\scriptsize red}}})}$, acts as a control parameter governing the transition between simulations with and without rebound, as predicted by Eq.~\eqref{eq:R0(A')}}.
        \label{fig:scatterplot}
    \end{figure*}
    ﻿
    In \cite{castioni_rebound_2024}, the authors revealed a critical feature of any finite vaccination campaign with a waning vaccine on the epidemiological dynamics of SIRS models. The effect is a ``rebound'' in the number of infected cases, caused by the synchronization of cases after waning begins to affect the population. A significant resurgence in cases is only possible if the number of susceptibles is allowed to build up above the critical threshold; this, in turn, occurs only if the vaccination campaign reduces the number of infections to a value close to zero. 
    ﻿
    As previously mentioned, in the mean-field case, this happens when $\alpha > \alpha_{\text{er}}$, but we are now interested in determining whether a similar threshold behavior exists concerning vaccination coverage (i.e., the number of vaccinated patches in the metapopulation).
     \\
    For instance, in the case of the network model described in Eqs.~\eqref{eq:SIRS1}-\eqref{eq:SIRS3}, the condition stated above is trivially satisfied if we vaccinate every single patch. However, it becomes less straightforward when only a fraction of patches are vaccinated, as this implies that we only have direct control over a subset of infections. 
    ﻿
    Interestingly, it turns out that the indirect control over the remainder of the network (which we will henceforth refer to as the \textit{reduced} network) can be quantified and exploited by once again using the reproduction number defined in Eq.~\eqref{eq:R0(A)}. Specifically, if the reproduction number of the \textit{reduced} network is below one, the epidemic will spontaneously converge to zero (given a sufficiently long campaign duration), thereby satisfying the condition for a rebound to occur once the vaccination is lifted.
    In mathematical terms, if we define $\mathbf{A_{\mbox{\scriptsize red}}}$ as the \textit{reduced} adjacency matrix obtained by removing all the patches that have been vaccinated, then the rebound happens only if 
    \begin{equation}\label{eq:R0(A')}
        R_0{\bf(A_{{\mbox{\scriptsize red}}})} = \dfrac{\beta}{\mu} (1 + p \lambda_{\mbox{\scriptsize max}}\bf(A_{{\mbox{\scriptsize red}}}) ) < 1
    \end{equation}
    A schematic explanation of this phenomenon is given by Fig. \ref{fig:scheme}. \\
    This condition provides critical insights into the effects of random vaccination strategies. For instance, vaccinating a small number of patches will not induce a rebound, as it will not significantly alter the spectral properties of the network, even when targeting a substantial fraction of the network's patches. This is due to the fact that, particularly in heterogeneous networks, the majority of patches are peripheral and play a negligible role in the overall dynamics.
    ﻿
    To confirm this intuition, we simulated 200 different vaccination strategies targeting 160 randomly chosen patches out of a 200 patch heterogeneous network. For each scenario, the reduced $R_0$ and the height of the rebound were recorded and plotted in a scatter plot shown in Fig.~\ref{fig:scatterplot}. The results reveal a clear distinction between strategies that induce a reproduction number below or above one. In the former case, vaccination leads to a significant rebound, while in the latter, only minor oscillations around the equilibrium value are observed.
    ﻿
    Furthermore, the figure demonstrates that all vaccination strategies including the highest-degree patches have a high probability of inducing a rebound. This observation is supported by basic network theory, specifically the Perron-Frobenius theorem, which states that $\lambda_{\max}(\mathbf{A}) \leq k_{\max}$, i.e., the maximum degree serves as an upper bound for the spectral radius of the adjacency matrix. Consequently, it follows that, especially in heterogeneous networks, targeting the highest-degree patches is consistently more effective at reducing the reproduction number. As a result, such targeted strategies lead to larger rebounds once vaccination efforts are lifted.
     \\
    ﻿
    Finally, we should emphasize that, for large vaccination coverages, the reduced network will likely become disconnected. One might think that this necessitates treating each component separately, calculating the reproduction number for each. However, this is not required, as Eq.~\eqref{eq:R0(A')} inherently provides the largest reproduction number among all components in the disconnected network. Thus, if we can verify that this value is below 1, we can automatically conclude that the same condition holds for all other components of the network as well.
    ﻿
    For this reason, throughout this work, we will refer to $R_0\left(\mathbf{A}_{\mbox{\scriptsize red}}\right)$---\textit{with a slight abuse of notation}---as the basic reproduction number of the \textit{entire} reduced network, rather than that of any specific component.
    ﻿
    ﻿
    \begin{figure*}[t]
        \centering
        \includegraphics[width=0.9\linewidth]{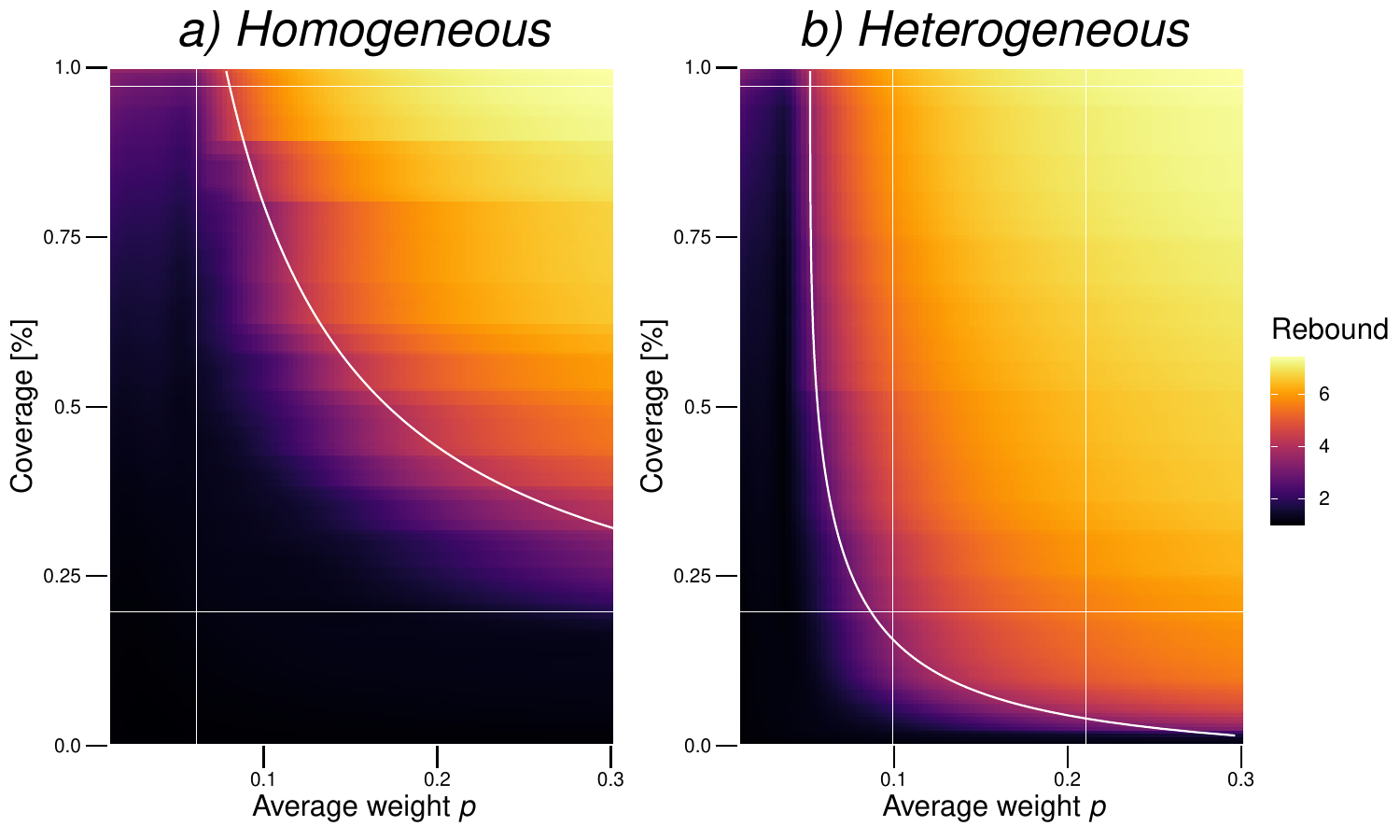}
     \caption{The heatmaps in this figure show the conditions under which the rebound appears, based on the average weight $p$ (x-axis) and the percentage of vaccine coverage (y-axis). The left and right panels correspond to a homogeneous network ($a$) and a heterogeneous network ($b$), respectively, both with $N=200$ and $\expval{k} \simeq 4$. The color scale represents the rebound height normalized by the stationary value of the $I$ compartment. The white line corresponds to the theoretical prediction of the critical line given by Eq.~\eqref{eq:p_crit}, obtained by calculating $p_{\text{crit}}$ for each coverage value from 1 to $N-1$ and smoothing it using a mathematical process described in the Appendix.}
        \label{fig:heatmap}
    \end{figure*}
    ﻿
    \subsection*{Targeted vaccination and critical mobility}
    ﻿
    Up to this point, we have focused on how the presence of a rebound is determined by the combination of patches chosen for vaccination, particularly in the context of random strategies. However, examining Eq.~\eqref{eq:R0(A')}, it becomes evident that the average link weight $p$ also plays a significant role in determining whether a rebound occurs. In this section, we explore the interplay between $p$ and the extent of vaccination coverage.
    ﻿
    To this end, we consider a mobility network with a fixed structure and an average link weight $p$ varying between 0.01 and 0.3. The vaccination coverage is allowed to range from 1 to $N-1$ (where $N$ is the size of the network). To avoid spurious effects arising from variations in the overall basic reproduction number---which itself depends on $p$, as per Eq.~\eqref{eq:R0(A)}---we rescale the transmission rate $\beta$ by making it a function of $p$:
    \begin{equation} \label{eq:beta_rescaled}
        \beta(p) = \dfrac{\beta_0}{1 + p \lambda_{\max}(\mathbf{A})},
    \end{equation}
    such that the reproduction number remains constant and independent of $p$, i.e., $R_0(\mathbf{A}) = \beta_0 / \mu$, where $\beta_0$ is the transmission rate in a totally disconnected patch.
    ﻿
    Finally, we restrict our analysis to targeted vaccination campaigns, where patches are vaccinated in decreasing order of their degree. This procedure is repeated for both a homogeneous and a heterogeneous network.
    \\
    ﻿
    Given these premises, it is possible to analytically derive the critical mobility, i.e., the value of $p$ above which, for a given set of vaccination targets, a rebound always occurs. This condition is obtained by substituting Eq.~\eqref{eq:beta_rescaled} into Eq.~\eqref{eq:R0(A')},setting the latter equal to 1, and solving for $p$. The resulting expression is:
    ﻿
    \begin{equation}\label{eq:p_crit}
        p_{\mbox{\scriptsize crit}} = \qty(\dfrac{\beta_0}{\mu} - 1) \dfrac{1}{\lambda_{\mbox{\scriptsize max}}(\mathbf{A}) - \frac{\beta_0}{\mu} \lambda_{\mbox{\scriptsize max}}(\mathbf{A}_{\mbox{\scriptsize red}})}.
    \end{equation}
    ﻿
    This formula not only identifies the location of the critical line but also provides two important threshold conditions. First, since $p$ must be a positive number, the equation becomes meaningless when the denominator is negative. This leads to the following threshold condition:
    ﻿
    \begin{equation}\label{eq:condition}
        \lambda_{\mbox{\scriptsize max}}(\mathbf{A}_{\mbox{\scriptsize red}}) < \dfrac{\mu}{\beta_0} \lambda_{\mbox{\scriptsize max}}(\mathbf{A}),
    \end{equation}
    ﻿
    which identifies the vaccination distributions where a rebound could potentially occur. 
    ﻿
    Second, the other possible limit of Eq.~\eqref{eq:p_crit} occurs when $\lambda_{\mbox{\scriptsize max}}(\mathbf{A}_{\mbox{\scriptsize red}}) = 0$, which corresponds to vaccinating all but one patch. In this case, we obtain the following threshold value for $p$:
    ﻿
    \begin{equation}
        p > \qty(\dfrac{\beta_0}{\mu} - 1) \dfrac{1}{\lambda_{\mbox{\scriptsize max}}(\mathbf{A})},
    \end{equation}
    ﻿
    representing the value of the weight $p$ below which no rebound is possible.
    ﻿
    Evidence for this effect is presented in the results shown in Fig.~\ref{fig:heatmap}. The most noticeable difference between the homogeneous and heterogeneous networks is that the former exhibits a wide range of vaccination coverages where a rebound does not manifest, while in the latter, this range is as narrow as just a few patches. This difference can be interpreted in light of a standard result from network theory, which states that the spectral radius of a network is proportional to the average degree $\expval{k}$ in homogeneous networks, while in heterogeneous networks, it is proportional to the square root of the maximum degree $k_{\mbox{\scriptsize max}}$ \cite{sarkar_spectral_2018}. Incorporating this insight into the condition given in Eq.~\eqref{eq:condition} clarifies the difference observed in the panels of Fig.~\ref{fig:heatmap}. In homogeneous networks, a significant number of patches must be removed before this impacts the average degree, due to the relative uniformity of the network. Conversely, in heterogeneous networks, the targeted removal of hubs can substantially reduce the spectral radius in just a few steps. 
    ﻿
    This same reasoning explains the difference in the accuracy of the critical line between the two panels. In heterogeneous networks, the rapid drop in $k_{\mbox{\scriptsize max}}$ causes an abrupt transition between rebound and non-rebound regimes. On the other hand, in homogeneous networks, the slow decline of $\expval{k}$ leads to a much smoother transition, which makes the critical line less precise but still qualitatively valid. 
    ﻿
    Finally, in both panels, the top-left corner is slightly lighter than the rest of the non-rebound region. This occurs because, when the vaccination coverage approaches 100\%, the number of infected individuals is non-negligible in only a few patches. As a result, the time required for the infection to spread throughout the entire network allows the susceptible population to build up slightly beyond its equilibrium value, fueling a minor resurgence of infections even if we are outside of the rebound regime.
    ﻿
    ﻿
    \begin{figure*}
        \centering
        \includegraphics[width=0.9\linewidth]{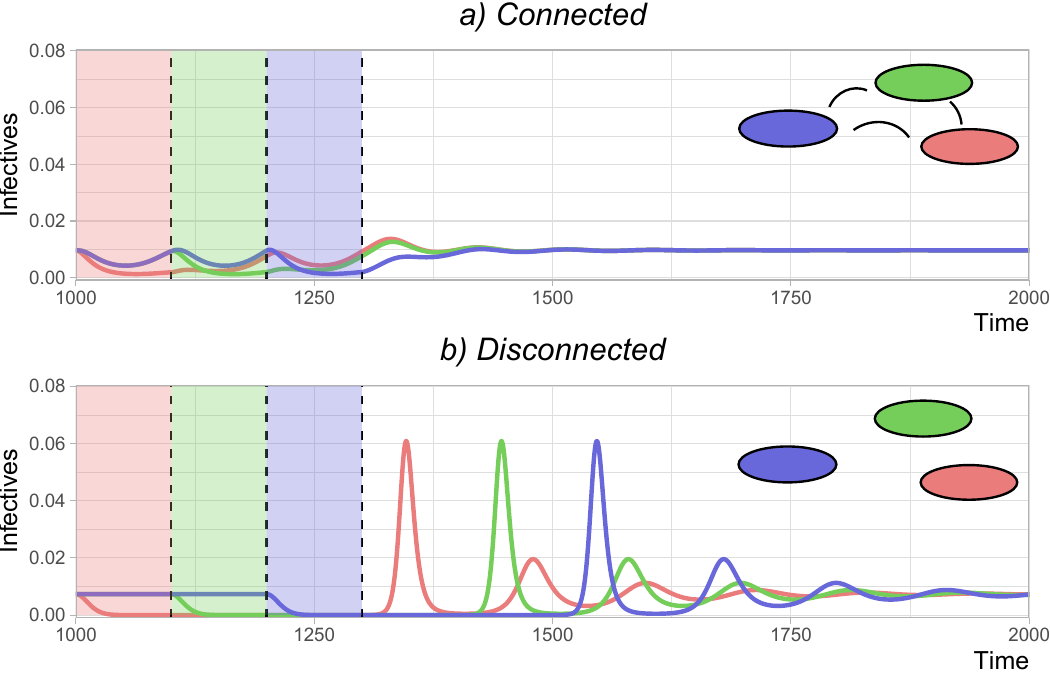}
        \caption{\textbf{Asynchronous vaccination in a connected ($a$) and disconnected ($b$) three-patches network.} . In both cases, the time starts from 1000 because the system was allowed to reach equilibrium before the vaccination began. The red, green, and blue shaded areas represent the time windows during which each of the three patches was vaccinated, as explained in Eq.~\eqref{eq:async}. It is evident that the asynchronous strategy significantly reduces the rebound in the connected case compared to the disconnected one. To ensure comparability between the two systems, they were assigned the same $R_0$ by adjusting $\beta(p)$ as prescribed by Eq.~\eqref{eq:beta_rescaled}. The parameters used were $\beta_0 = 0.8$, $\mu = 0.5$, $\delta = 0.01$, and $\alpha = 0.01$. Finally, $p$ was set to $0.3$ for the connected network and to $0$ for the disconnected one.}
        \label{fig:async}
    \end{figure*}
    ﻿
    \subsection*{Asynchronous vaccination}
\begin{figure}
        \centering
        \includegraphics[width=0.8\linewidth]{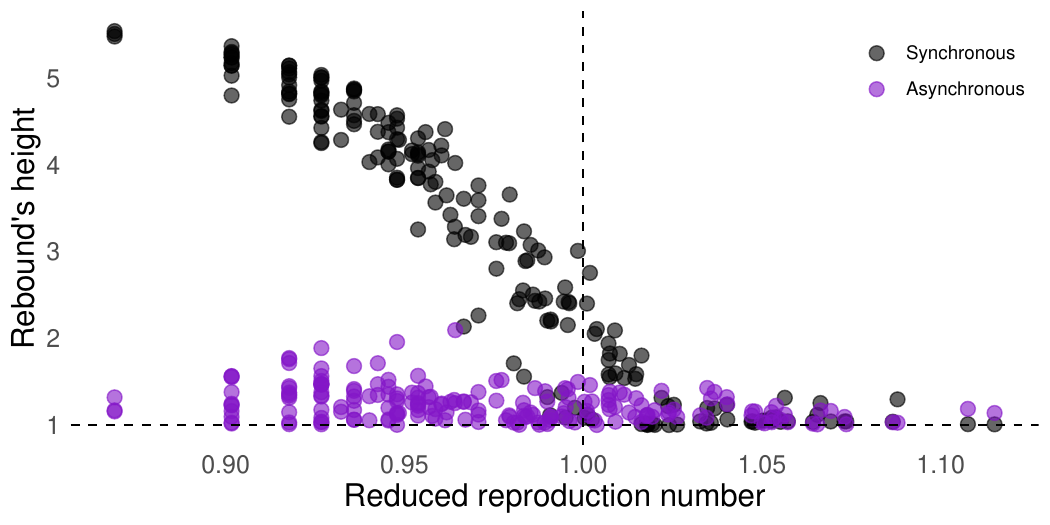}
        \caption{{\bfseries Effects of asynchronous vaccinations}. In this figure, each point represents a different realization of a vaccination campaign. The black points correspond to those plotted in Fig. \ref{fig:scatterplot}, where vaccination for all targeted nodes began and ended during the same time period. The purple points, on the other hand, represent the same spatial distribution of vaccines but with an asynchronous roll-out, in which the total number of targeted nodes was divided into five groups that received their doses one after another. The difference between the two groups is striking, highlighting that the rebound effect disappears when asynchronous vaccinations are employed.}
        \label{fig:scatterplot_async}
    \end{figure}

    We understand the rebound as an effect caused by the synchronized disappearance of infection from several patches at the same time. Therefore, it is not the sheer amount of vaccines that causes it, but rather the timing of their administration. Using this insight, we propose an alternative vaccination strategy where the same amount of vaccines is distributed asynchronously, i.e., with different starting times for each patch. This approach ensures that all doses are used while preventing the number of infected individuals from approaching zero.
    ﻿
    To test this idea, we propose the following strategy: we took a three-patch network and simulated an epidemic with the usual SIRS model in Eq.~\eqref{eq:SIRS1}-\eqref{eq:SIRS3} until it reached the stationary state. Then, we started a vaccination campaign targeting one patch at a time:
    ﻿
    \begin{equation} \label{eq:async}
        \alpha_i^{\mbox{\scriptsize async}} = \begin{cases}
            \alpha \delta_{1i} \quad \text{if $t\in \qty[ t_{\mbox{\scriptsize start}}, t_2 ]$} \\
            \alpha \delta_{2i} \quad \text{if $t\in \qty[ t_2, t_3 ]$} \\
            \alpha \delta_{3i} \quad \text{if $t\in \qty[ t_3, t_{\mbox{\scriptsize stop}} ]$}  \\
            0 \quad \text{otherwise}
        \end{cases}
    \end{equation}
    Finally we compare the resulting rebound with the once obtained with the same strategy but in the case of purely isolated patched, without network structure but with the same reproduction number. The result of this comparison are shown in Fig. \ref{fig:async}, where we can see that a startling difference between the connected case, where by making sure that the infected do not drop too close to zero the rebound practically disappears, while in the case of isolated patches the rebounds happens not one but three times, one for each patch. \\
    ﻿
    In Fig. \ref{fig:scatterplot_async} we also show that the same idea works on every single strategy displayed in Fig. \ref{fig:scatterplot}, when instead of vaccinating all the patches at the same time we split them into five groups and vaccinate them one after the other. It that way it can be seen that the height of the rebound sharply drops even when the reduced reproduction number is below 1.

    ﻿
    ﻿
    ﻿
    \section*{Discussion}
    The literature on epidemic control, especially the one regarding vaccination, has always been overwhelmingly focused on analyzing the impact of different types of vaccination campaign in terms of some established measures such as the height of the peak, its position or the attack rate. Significant resources have been dedicated to identifying “optimal” vaccination strategies—those that minimize these metrics. However, less attention has been paid to the dynamics that unfold after such campaigns end, whether due to design or resource constraints. This gap in the literature is critical, as it overlooks the long-term consequences of vaccination efforts, particularly in systems where reinfections or other mechanisms replenish the susceptible pool. In this study, we investigated what happens when an epidemic control policy temporarily suppresses the number of infected individuals without eliminating them entirely. Our findings demonstrate that such policies create conditions for a subsequent resurgence, known as the ``rebound effect''. This effect arises because the artificially low prevalence during the campaign allows the susceptible population to accumulate, setting the stage for a sharp resurgence once the campaign ceases. Using a metapopulation model, we have shown that this rebound phenomenon is a robust feature of epidemic models with reinfections or similar mechanisms, independent of the specific details of the contagion process.\\
    After establishing the existence of the rebound effect in networks with complex topologies, we identified the minimum requirements for its occurrence. Our analytical criterion, summarized in Eq. \eqref{eq:condition}, states that a rebound will occur if the fraction of the network left unvaccinated has a reproduction number below one. 
    ﻿
    Paradoxically, this leads to the counterintuitive conclusion that the most effective vaccination strategies—those that reduce the basic reproduction number ($R_0$) the most—are also those most likely to set the stage for a severe rebound. For example, in the context of the ``city and villages'' model (where one big patch is assumed to be surrounded by a number of smaller ones), the standard strategy would be to target the city and let the villages reap the benefits indirectly. In the context of our model however, this would invariably lead to a huge resurgence, for the reasons stated above (see Fig. \ref{fig:scheme}). \\
    It is not our intention to suggest that there are situations in which vaccines should be avoided altogether. On the contrary, a precise understanding of the rebound effect has led us to identify asynchronous vaccination strategies as a solution to the problem, while keeping the number of vaccines deployed unchanged. This approach is so effective that it might explain why the rebound effect is rarely observed in real-world data: natural variations in the starting times of vaccination strategies may smooth the effect to the extent that it becomes indistinguishable from noise.\\ 
    ﻿
    ﻿
    \section*{Acknowledgements}
    \noindent Work supported by Spanish Ministerio de Ciencia e Innovaci\'on (PID2021-128005NB-C21), Generalitat de Catalunya (2021SGR-00633), Universitat Rovira i Virgili (2023PFR-URV-00633) and the European Union’s Horizon Europe Programme under the CREXDATA project (grant agreement no.\ 101092749). AA acknowledges ICREA Academia, the James S.\ McDonnell Foundation (Grant N.\ 220020325), and the Joint Appointment Program at Pacific Northwest National Laboratory (PNNL). PNNL is a multi-program national laboratory operated for the U.S.\ Department of Energy (DOE) by Battelle Memorial Institute under Contract No.\ DE-AC05-76RL01830.
    ﻿
    ﻿
\section*{Appendix A}
In Fig. \ref{fig:heatmap} of the main text the white critical line is presented as the solution of the equation:

\begin{equation}\label{eqsm:p_crit}
    p_{\mbox{\scriptsize crit}} = \qty(\dfrac{\beta_0}{\mu} - 1) \dfrac{1}{\lambda_{\mbox{\scriptsize max}}(\mathbf{A}) - \frac{\beta_0}{\mu} \lambda_{\mbox{\scriptsize max}}(\mathbf{A}_{\mbox{\scriptsize red}})}.
\end{equation}

\begin{figure}
    \centering
    \includegraphics[width=0.9\linewidth]{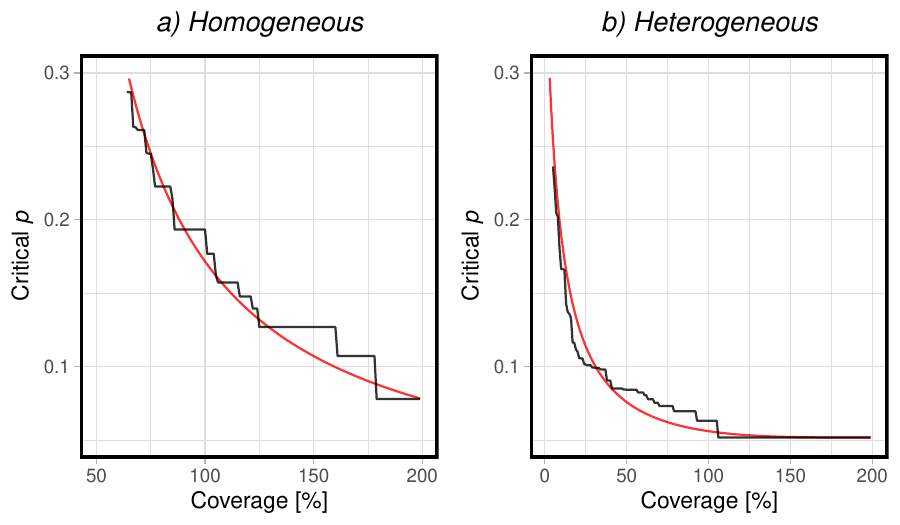}
    \caption{Comparison between the simple solution of Eq.~\ref{eqsm:p_crit} (black line) and the smoothed version (red line) obtained using the method described in the Appendix. The $x$-axis represents the coverage, measured as the absolute number of vaccinated nodes, while the $y$-axis shows the critical value of $p$ for the corresponding vaccination coverage.}
    \label{fig:smooth}
\end{figure}

In particular, this curve was calculated by sequentially removing nodes from the network, one at a time, in order of decreasing degree. For each removal, we calculated $\lambda_{\mbox{\scriptsize max}}(\mathbf{A}_{\mbox{\scriptsize red}})$, the spectral radius of the reduced adjacency matrix. This process produced a set of $p_{\mbox{\scriptsize crit}}$ values, each corresponding to a specific vaccination coverage. However, the curve obtained in this manner exhibits a step-like appearance due to the discrete nature of the node removal process. To solve the problem we fitted such step-like curve to the easiest possible function, which was:

\begin{equation}
    \lambda_{\mbox{\scriptsize max}}(\mathbf{A}_{\mbox{\scriptsize red}}) = A \dfrac{N_{\mbox{\scriptsize red}}}{N},
\end{equation}
for the homogeneous network and

\begin{equation}
    \lambda_{\mbox{\scriptsize max}}(\mathbf{A}_{\mbox{\scriptsize red}}) = A \qty(\dfrac{N_{\mbox{\scriptsize red}}}{N}
    )^\gamma,
\end{equation}
for the heterogeneous network, where $A$ and $\gamma$ are both free parameters determined by calibration. In particular the values we used were $A=4$ for homogeneous case and $A = 4.8$ and $\gamma = 3.5$ for the heterogeneous one. See Fig. \ref{fig:smooth} for a comparison with the smoothed and non-smoothed solution.

    ﻿
    ﻿

\bibliographystyle{unsrt}

\begin{thebibliography}{10}


\bibitem{bernoulli_attempt_2004}
Daniel Bernoulli and Sally Blower.
\newblock An attempt at a new analysis of the mortality caused by smallpox and
  of the advantages of inoculation to prevent it.
\newblock {\em Reviews in Medical Virology}, 14(5):275--288, September 2004.


\bibitem{klepac_synthesizing_2011}
Petra Klepac, Ramanan Laxminarayan, and Bryan~T. Grenfell.
\newblock Synthesizing epidemiological and economic optima for control of
  immunizing infections.
\newblock {\em Proceedings of the National Academy of Sciences {USA}},
  108(34):14366--14370, 2011.


\bibitem{lauro_optimal_2021}
Francesco~Di Lauro, István~Z. Kiss, and Joel~C. Miller.
\newblock Optimal timing of one-shot interventions for epidemic control.
\newblock {\em PLOS Computational Biology}, 17(3):e1008763, 2021.


\bibitem{wallinga_optimizing_2010}
Jacco Wallinga, Michiel van Boven, and Marc Lipsitch.
\newblock Optimizing infectious disease interventions during an emerging
  epidemic.
\newblock {\em Proceedings of the National Academy of Sciences {USA}},
  107(2):923--928, 2010.


\bibitem{morris_optimal_2021}
Dylan~H. Morris, Fernando~W. Rossine, Joshua~B. Plotkin, and Simon~A. Levin.
\newblock Optimal, near-optimal, and robust epidemic control.
\newblock {\em Communications Physics}, 4(1):1--8, 2021.


\bibitem{bubar_21}
Kate~M. Bubar, Kyle Reinholt, Stephen~M. Kissler, Marc Lipsitch, Sarah Cobey,
  Yonatan~H. Grad, and Daniel~B. Larremore.
\newblock Model-informed covid-19 vaccine prioritization strategies by age and
  serostatus.
\newblock {\em Science}, 371(6532):916--921, 2021.


\bibitem{pulsating_20}
Benjamin Steinegger, Alex Arenas, Jes\'us G\'omez-Garde\~nes, and Clara
  Granell.
\newblock Pulsating campaigns of human prophylaxis driven by risk perception
  palliate oscillations of direct contact transmitted diseases.
\newblock {\em Phys. Rev. Res.}, 2:023181, May 2020.


\bibitem{castioni_rebound_2024}
Piergiorgio Castioni, Sergio Gómez, Clara Granell, and Alex Arenas.
\newblock Rebound in epidemic control: how misaligned vaccination timing
  amplifies infection peaks.
\newblock {\em npj Complexity}, 1(1):20, November 2024.


\bibitem{behavioural_22}
Benjamin Steinegger, Lluís Arola-Fernández, Clara Granell, Jesús
  Gómez-Gardeñes, and Alex Arenas.
\newblock Behavioural response to heterogeneous severity of covid-19 explains
  temporal variation of cases among different age groups.
\newblock {\em Philosophical Transactions of the Royal Society A: Mathematical,
  Physical and Engineering Sciences}, 380(2214):20210119, 2022.


\bibitem{protective_22}
Mozhgan Khanjanianpak, Nahid Azimi-Tafreshi, Alex Arenas, and Jesús
  Gómez-Gardeñes.
\newblock Emergence of protective behaviour under different risk perceptions to
  disease spreading.
\newblock {\em Philosophical Transactions of the Royal Society A: Mathematical,
  Physical and Engineering Sciences}, 380(2227):20200412, 2022.


\bibitem{mclean_measles_1988}
A.~R. McLean and R.~M. Anderson.
\newblock Measles in developing countries. {Part} {II}. {The} predicted impact
  of mass vaccination.
\newblock {\em Epidemiology and Infection}, 100(3):419--442, June 1988.


\bibitem{honey_21}
N.~Akhavan Kharazian and F.~M.~G. Magpantay.
\newblock The honeymoon period after mass vaccination.
\newblock {\em Mathematical Biosciences and Engineering}, 18(1):354--372, 2021.


\bibitem{anderson_vaccination_1983}
R.~M. Anderson and R.~M. May.
\newblock Vaccination against rubella and measles: quantitative investigations
  of different policies.
\newblock {\em The Journal of Hygiene}, 90(2):259--325, April 1983.


\bibitem{mclean_after_1995}
AngelaR McLean.
\newblock After the honeymoon in measles control.
\newblock {\em The Lancet}, 345(8945):272, February 1995.


\bibitem{scherer_mathematical_2002}
Almut Scherer and Angela McLean.
\newblock Mathematical models of vaccination.
\newblock {\em British Medical Bulletin}, 62(1):187--199, July 2002.


\bibitem{bhattacharyya_age-specific_2017}
S.~Bhattacharyya and M.~J. Ferrari.
\newblock Age-specific mixing generates transient outbreak risk following
  critical-level vaccination.
\newblock {\em Epidemiology \& Infection}, 145(1):12--22, January 2017.


\bibitem{munday_estimating_2024}
James~D. Munday, Katherine~E. Atkins, Don Klinkenberg, Marc Meurs, Erik Fleur,
  Susan~JM Hahné, Jacco Wallinga, and Albert Jan~van Hoek.
\newblock Estimating the risk and spatial spread of measles in populations with
  high {MMR} uptake: {Using} school-household networks to understand the 2013
  to 2014 outbreak in the {Netherlands}.
\newblock {\em PLOS Medicine}, 21(10), 2024.


\bibitem{may_spatial_1984}
Robert~M. May and Roy~M. Anderson.
\newblock Spatial heterogeneity and the design of immunization programs.
\newblock {\em Mathematical Biosciences}, 72(1):83--111, November 1984.


\bibitem{sattenspiel_geographic_2009}
Lisa Sattenspiel.
\newblock {\em The {Geographic} {Spread} of {Infectious} {Diseases}: {Models}
  and {Applications}: {Models} and {Applications}}.
\newblock Princeton University Press, December 2009.


\bibitem{granell_24}
Clara Granell, Sergio Gómez, Jesús Gómez-Gardeñes, and Alex Arenas.
\newblock Probabilistic discrete-time models for spreading processes in complex
  networks: A review.
\newblock {\em Annalen der Physik}, 536(10):2400078, 2024.


\bibitem{sarkar_spectral_2018}
Camellia Sarkar and Sarika Jalan.
\newblock {Spectral properties of complex networks}.
\newblock {\em Chaos: An Interdisciplinary Journal of Nonlinear Science},
  28(10):102101, 10 2018.


\end{thebibliography}

\end{document}